\documentstyle[11pt]{article}
\input{epsf}

\def\1{\mbox{I\hspace{-.15em}1}}

\def\b{\begin{equation}}
\def\e{\end{equation}}
\def\bee{\begin{enumerate}}
\def\eee{\end{enumerate}}

\title{Gravitational wave detection by bounded \\cold electronic plasma in a long pipe}
\author{O. Jalili\thanks{E-Mail: omid\_jalili@yahoo.com }, S. Rouhani and M.V.
Takook\thanks{E-Mail: takook@razi.ac.ir}}
\date{\today}

\begin{document}
  \maketitle {\centerline{\it
Department of Physics, Science and Research Branch, Islamic Azad University, }
 \centerline{\it P.O.BOX 14835-157, Tehran, IRAN}\it }
\begin{abstract}

We intend to propose an experimental sketch to detect
gravitational waves (GW) directly, using an cold electronic plasma
in a long pipe. By considering an cold electronic plasma in a long
pipe, the Maxwell equations in 3+1 formalism will be invoked to
relate gravitational waves to the perturbations of plasma
particles. It will be shown that the impact of GW on cold
electronic plasma causes disturbances on the pathes of the
electrons. Those electrons that absorb energy from GW will pass
through the potential barrier at the end of the pipe. Therefore,
crossing of some electrons over the barrier will imply the
existence of the GW.
\end{abstract}

\vspace{0.5cm} {\it Proposed PACS numbers}: 04.62.+v, 98.80.Cq,
12.10.Dm \vspace{0.5cm}

\section{Introduction}

Stars and spaces between them are composed of matter in plasma
phase. Since gravitational field is the most effective force in
the macroscopic scale it will determine the dynamics of
astrophysical objects. So it is necessary to study the effects of
gravity on plasma phase of matter. For instance, the major reason
for spiral shape of galactic arms has been known to be the effect
of GW at the first stages of forming spiral galaxies \cite{Pad}.
Thus the theory of plasma state of matter in curved spacetime
has been carried out
\cite{Klaus,Betschart1,Servin2,subramanian,Meier,zunkel,Moortgat1,Marklund3,Brodin2,Brodin1,Marklund1}.
An interesting phenomenon that happens is creation of electromagnetic field by GW.
Gravitational fields enters into Maxwell Eqs as current terms, and like the electrical current
create a electromagnetic field \cite{Marklund1}. An interesting point in such studies is the emergence of resonance phenomenon in the interaction of GW  with plasma\cite{Servin3}. This fact is used to propose a new method for detecting GW, beside standard methods\cite{Weber,Gertsenshtein}.\\
As we know gravitational waves have not been detected directly in
the laboratory. The very small value of gravitational constant in
comparison with Coulomb constant causes such failures in detection
of GW in laboratory efforts. There have been many efforts to
suggest experiments in order to detect GW. Metal rod resonance and
optical interferometry have been the most common methods in
efforts to observe GW \cite{Weber,Gertsenshtein}. There are some
other methods too \cite{Chiao1}. The most successful method
implying the existence of GW has been the energy reduction of
binary stars \cite{Taylor1,Taylor2}. In fact we have never
observed the gravitational waves. Taylor has computed the energy
of gravitational radiation of binary stars rotating about each
other. Observed energy reduction of such stars is in full
agreement with Taylor calculation.

The most promising idea is that of Weber. To observe GW effects
directly in the laboratory, Weber has computed the effect of GW on
the length of an Aluminum bar. We have used the basic Weber's idea
but changed some ingredients. We thought that replacing Weber's
rod with an electronic bar may be useful. Actually electronic bar
has a great advantage; the electronic bar does not have the
unwanted solid structure and then, atomic structures does not have
noisy effects on it. Eliminating noisy effects is very important
because they may interfere with GW. Our main purpose is to
separate GW from environmental effects as much as possible. In
order to do so, we enclose a cold electronic plasma in an
potential well First, figure 1 and then two negative rings will be
placed at the ends of the pipe to create a potential barrier. The
potential barrier will enclose electronic cloud within the
cylinder. It can be shown that the slope of this barrier is very
sharp, so the trapped electrons will encounter to the potential
barrier very close to the two rings \cite{Rouhani}. The rings will
create  radial fields that will be discussed later. We will place
a shield on the whole apparatus to prevent it from environmental
noisy effect. Thus there is no any external electromagnetic field
affecting on the pipe. But GW will pass through the metallic wall
and will affect on the dynamics of the internal electrons. Length
of the pipe should be suitably chosen \cite{Chiao2}. Electrons
within the pipe will get energy from GW and will eventually pass
through the ring's potential barrier.
Almost all electron are trapped in the pipe but small number
of electrons can cross the barrier due to energy absorbtion from GW.
The escaped electrons may be simply detected by an electronic detector.

In this paper we consider first bounded cold electronic plasma in a long pipe.
The effect of Perturbation in electronic motion will be considered in
section 3. Then we review 3+1 formalism to explain why GW has real existence.
We will consider in section 5 the effect of GW on
cold electronic plasma that is enclosed in a long pipe.
Finally we propose this apparatus for a possible detector of GW.

\section{Enclosed cold electronic plasma in a long pipe}

We have previously studied the electric field inside a long pipe
equipped two negative charge rings and shown that it is damped as
cosine hyperbolic function \cite{Rouhani}. Actually there can not
be any sign of electric field created by the rings within the
pipe. Suppose an amount of cold  electronic cloud within the pipe
is imported. We impose a longitudinal magnetic field $B_{0}$ to
radially enclosed electronic cloud, \textbf{Fig 1}. The electrons
will rotate around the pipe axis. During rotation, they will move
to the ends of  the pipe. At the ends, the electrons will confront
the barrier and will be reflected. We will use the fluid model to
describe the situation, so there will not be pure longitudinal
motion and any $B_{\varphi}$ due to the longitudinal motion of
electrons. Rotational motion of the electrons will create a
$B_{z}$ that will oppose the $B_{0}$ (Lens law). There is some
$E_{r}$ due to non-neutral nature of the electronic plasma. We
will now compute this $E_{r}$ field from Poisson equation:

\begin{figure}[tb]
  \centerline{\epsfxsize=4.5in \epsfysize=1.5in{\epsffile{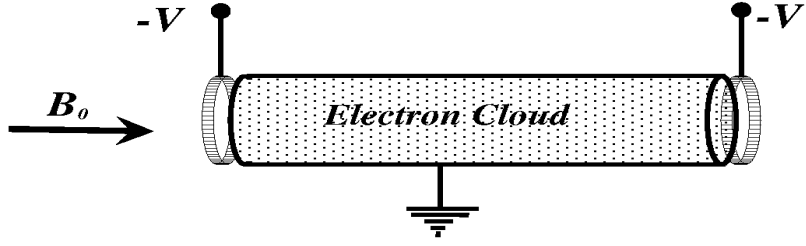}}}
    \caption{Enclosed cold electronic plasma in a long pipe. Two negative potential rings at the end of the pipe create a potential barrier.}\label{11}
\end{figure}

\b\vec \nabla.\vec E=-4\pi e n^{0} \Rightarrow
\frac{1}{r}\frac{\partial}{\partial r}(rE_{r})
     =-4\pi e n^{0}\Rightarrow E_{r}=-2\pi e n^{0}r.\e
To obtain $B_{z}$, we use the Steady state Ampere law: \b \vec
\nabla\times \vec B=\frac{4\pi}{c}(-n^{0}e \vec v).  \e Since the
vector $\vec v$ merely have $v_{\varphi}$ component, we conclude
from above equation that \b \frac{\partial B^0_z}{\partial r}
=\frac{4\pi}{c}n^0ev_\phi. \e Thus longitudinal magnetic field
will be changed with radius by the rotational electron motion and
then, longitudinal magnetic field will never be a uniform field.
We will begin with momentum equation to find rotational frequency
of the electrons around axis. The plasma is so cold that there is
no pressure or impact terms in momentum equation. Thus: \b
nm(\frac{\partial }{\partial t}+\vec v.\vec \nabla)v=-en^0(\vec
E+\frac{\vec v\times \vec B}{c}), \e where $n^{0}$ is the
electrons density, $m$ the electron mass, and $e$ the electron
charge. Electric field in the cylinder is  \b \vec
E=E_r\overrightarrow{a}_r+E_z\overrightarrow{a}_z .\e $E_{z}$ is
largely due to lateral potential rings and a little due to
finiteness of electronic column. As we have quoted earlier
\cite{Rouhani}, the $E_{z}$ of the rings will rapidly be damped,
thus $E_{z}$ will only be non-vanishing closely near (in some
millimeter) the ends. We will ignore the $E_{z}$ that belongs to
the finiteness of the electronic column. $E_{r}$ can be obtained
from Eq (2). The $\vec B$ field has two components: \b \vec
B=B_\phi\overrightarrow{a}_\phi+B_z\overrightarrow{a}_z .\e
$B_{z}$ is due to longitudinal motion that will be ignored in the
fluid model, because the longitudinal current average is zero.
$B_{z}$ has two components too: \b B_z=B^0+B^0_z,  \e where
$B^{0}$ is the confining longitudinal external field and
$B^{0}_{z}$ is due to the rotational motion. $B^{0}_{z}$ can be
ignored in comparison with $B^{0}$. We then have from Eq(4): \b
\frac{1}{r}(v_\phi)^2(-\overrightarrow{a}_r)=
\frac{-e}{m}(E_r+\frac{v_\phi B^0}{c})\overrightarrow{a}_r. \e And
then: \b\omega _r=\frac{v_\phi^0}{r} =\frac{\omega _c}{2}[1\pm(1-
\frac{2\omega_p^2}{\omega_c^2})^\frac{1}{2}] ,\e where $
\omega_c=\frac{eB_0}{mc}$  is the cyclotron angular frequency of
pseudo-neutral plasma and $\omega_p=\sqrt{\frac{4\pi e^2}{m}}$ is
the angular frequency of the electronic plasma. If the plasma was
not pseudo-neutral then there would not be any $E_r$. So we can
conclude from Eq(4): $\omega_r=\omega_c$. It is obvious from Eq(9)
that if we did not ignore $B^{0}_{z}$ in comparison with $B^{0}$
then $B_z$ would be changed with radius (Eq(3)). If doing so,
$\omega_c$ will vary with $r$ too, and then $\omega_r$ vary with
$r$. But the following calculation shows that the variation is so
small: \b \frac{\partial B^0}{\partial r}  =4\pi n^0 e v_\phi^0
 =\frac{4\pi n^0 e}{c}\frac{\omega_c}{2}[1\pm(1-\frac{2\omega_p^2}{\omega_c^2})^\frac{1}{2}]r
 =\frac{2\pi n^0 e^2 B^0}{mc^2}[1\pm(1-\frac{2\omega_p^2}{\omega_c^2})^\frac{1}{2}]r\ll1. \e

\section{The effect of perturbation on the dynamic of electrons}

Maxwell's Eqs and Newton's Eqs dominate on perturbations namely
perturbation can not get arbitrary values. Suppose an external
agitations cause a  first order perturbation: \b n=n_0+n_1, \; \;
\; \;\; \; \; \;
      \vec B=\vec B_0+\vec B_1,
\; \; \; \;\; \; \; \;  \vec v=\vec v_0+\vec v_1,
     \; \; \; \;\; \; \; \; \vec E=\vec E_1.\e
We ignored the constant part of $E$. Starting by the first order
momentum equation, one finds: \b \frac{\partial \vec v_1}{\partial
t}+\vec v_0.\vec \nabla \vec v_1+\vec v_1.\vec \nabla \vec
v_0=\frac{-e}{m}(\vec E_1+\frac{\vec v_0\times \vec B_1+\vec
v_1\times \vec B_0}{c}). \e Suppose the perturbation merely has
radial and polar components: \b\vec v^1=v_\phi
\overrightarrow{a}_\phi+v_r \overrightarrow{a}_r , \e where it is
a mode. There are many electromagnetic modes that correspond to
excitant fields. We will see that in the first order, the shear
tensor can only exist. Then only deformation in x and y components
will exist. Expressing Eq(2) in cylindrical coordinates:
\b\frac{\partial v_r}{\partial t}+\omega_r ( \frac{\partial
v_r}{\partial \phi}-v_\phi)-\omega_r v_\phi=\frac{-e}{m}E_r
-\frac{e}{mc}(v_0B_z+v_\phi B_0) ,\e \b   \frac{\partial
v_\phi}{\partial t}+\omega_r ( \frac{\partial v_\phi}{\partial
\phi}+v_r)
 +v_r\frac{\partial v_0}{\partial r}=\frac{-e}{m}E_\phi +\omega_c v_r,\e
We can express the above equations in Fourier domain
$\frac{\partial}{\partial t}=-i\omega$ and
$\frac{\partial}{\partial \varphi}=ik$, and using $\frac{\partial
v_0}{\partial r}=\omega_r$ approximation to obtain,
 \b  i(k\omega_r -\omega)v_r -\omega_r v_\phi =\frac{-e}{m}E_r -\frac{ev_0}{mc}B_z,\e
\b \omega_r v_r +i(k\omega_r-\omega)v_\phi=\frac{-e}{m}E_\phi. \e

If we solve this algebraic equations, we can express velocity in
term of electric and magnetic field components \b v_\phi =aE_r
+brB_z+fE_\phi,  \;\; v_r =AE_\phi +DE_r +SrB_z , \e where $
a,b,f,A,D,S$ are defined as follows \b a=\frac{-e\omega_r
/m}{(k\omega_r -\omega)^2 -\omega_r ^2},\;\;
b=\frac{-e\omega_r^2/mc}{(k\omega_r- \omega)^2
-\omega_r^2},\;\;f=\frac{i(k\omega_r -\omega)/m}{(k\omega_r-
\omega)^2-\omega_r^2}
$$ $$  A=\frac{-e}{m\omega_r}+\frac{(k\omega_r-
\omega)^2}{\omega_r m[(k\omega_r- \omega)^2 -\omega_r^2]},\;\;
D=\frac{i(k\omega_r -\omega)e/m}{(k\omega_r-
\omega)^2-\omega_r^2}=ef, \;\; S=\frac{ie\omega_r(k\omega_r
-\omega)/mc}{(k\omega_r- \omega)^2-\omega_r^2}. \e We need
electron density and velocity to compute electromagnetic's field
in the cylinder. Eqs(18) shows that velocity can be expressed in
terms of fields. Now we write density in terms of velocity. Since
unperturbed velocity has only $\phi$ component, we have from
linearized momentum equations; \b\frac{\partial n_1}{\partial
t}+n_0\vec \nabla .\vec v_1+\omega_r\frac{\partial n_1}{\partial
\phi} =0 .\e Converting into Fourier domain gives:
 \b -i\omega n_1 +n_0 \vec \nabla .\vec v_1 +ik\omega_r n_1 =0 \Longrightarrow n_1=\frac{n_0 \vec \nabla .\vec v_1}
 {i(\omega -k\omega_r)}. \e

Now, we begin with linearized Ampere Eqs and replace electron's
velocity and density from Eq(18) and Eq(21). First order Ampere
Eqs are \b \vec \nabla \times \vec B_1 =\frac{1}{c}\frac{\partial
\vec E^1}{\partial t}-\frac{4\pi e}{c}(n_0 \vec v_1 +n_1 \vec
v_0). \e Expressing Eqs(22) in cylindrical coordinates and then in
Fourier domain: \b \frac{i k}{r}B_z-ik'B_\phi
=\frac{-i\omega}{c}E_r- \frac{4\pi e n_0}{c}v_r ,$$ $$
\frac{-\partial B_z}{\partial r}=\frac{-i\omega}{c}E_\phi
-\frac{4\pi e n_0}{c}v_\phi - \frac{4\pi e v_0}{c}n_1, $$ $$
\frac{\partial}{\partial r}(rB_\phi)=0. \e Now, we put $n$ and $v$
from Eq(18) and Eq(21) in Eqs(23):
 \b \frac{ik}{r}B_z -ik'B_\phi =\frac{-i\omega}{c}E_r +(A'E_\phi +D'E_r +S'rB_z),  $$
 $$\frac{-\partial B_z}{\partial r}=\frac{-i\omega}{c}E_\phi
  +(a'E_r +b'rB_z +f'E_\phi ) +[A''\frac{\partial}{\partial r}(rE_\phi)+D''\frac{\partial}{\partial r}(rE_r)
  +s''\frac{\partial}{\partial r}(r^2 B_z)] , \e where
\b A'=\frac{-4\pi e n_0}{c}A ,\;\; D'=\frac{-4\pi e n_0}{c}D, \;\;
S'=\frac{-4\pi e n_0}{c}S ,$$ $$a'=\frac{-4\pi e
n_0}{c}(1+\frac{k\omega_r}{\omega-k\omega_r})a
,\;\;A''=\frac{-4\pi e\omega_r n_0}{ic(\omega-k\omega_r)}A .\e We
must eliminate one of $E$ or $B$ to solve Eq(24). Using Faraday
law in cylindrical coordinates, we can write $B$ in term of $E$ \b
B_\phi=\frac{ck'}{\omega}E_r,\;\;
B_z=\frac{c}{ir\omega}(\frac{\partial}{\partial
r}(rE_\phi)-ikE_r).\e Putting Eq(26) in Eq(24) and solving
$E_\phi$, gives: \b  (\frac{-i\omega}{c}+f')E_\phi+A''
\frac{\partial}{\partial r}(rE_\phi)
 +a'M(r)\frac{\partial}{\partial r}(rE_\phi)-a'N(r)E_\phi
 =
 \frac{-c}{i\omega}\frac{\partial}{\partial r}(\frac{1}{r}\frac{\partial}{\partial r}(rE_\phi))
 +\frac{kc}{\omega}\frac{\partial}{\partial r}(\frac{M(r)}{r}\frac{\partial}{\partial r}(rE_\phi))$$ $$
  -\frac{kc}{\omega}\frac{\partial}{\partial r}(\frac{N(r)}{r}E_\phi)
  -\frac{b'c}{i\omega}\frac{\partial}{\partial r}(rE_\phi)
  +\frac{k b'c M(r)}{\omega}\frac{\partial}{\partial r}(rE_\phi)
  -\frac{k b'c N(r)}{\omega}E_\phi
  -\frac{-S''c}{i\omega}\frac{\partial}{\partial r}(r\frac{\partial}{\partial r}(rE_\phi)),\e where
\b
M(r)=\frac{(\frac{ik}{r}-S'r)(\frac{c}{ir\omega})}{(\frac{-i\omega}{c}+D')
+ik(\frac{ik}{r}-S'r)(\frac{c}{ir\omega})+\frac{ik'^2
c}{\omega}},\;\;N(r)=\frac{A'}{(\frac{-i\omega}{c}+D')+ik(\frac{ik}{r}-S'r)(\frac{c}{ir\omega})+\frac{ik'^2
c}{\omega}} . \e

Solving Eq(27) and putting the field equal to zero at $r=R$, we
can obtain dispersion relation. Homogenous second order
differential Eq(27) has two solutions:
 \b E_{\phi}(r)=c_{1}E_{1\phi}+c_{2}E_{2\phi}.\e
One of this independent solution, for example $E_{2\phi}$, is
singular at origin and then is not physical. Reminder field must
satisfy the following boundary condition:
 \b E_{1\phi}(r=a)=0.\e By solving the obtained algebraic equation, one should obtain
desired dispersion relation: \b k=k(\omega).\e With series method
we have solved Eq(27) by MAPLE software. Based on what has been
said before, we have obtained dispersion relation. The result is
sketched in \textbf{Fig 2}.
\begin{figure}[hbt]
  \centerline{\epsfxsize=3.5in \epsfysize=3.5in{\epsffile{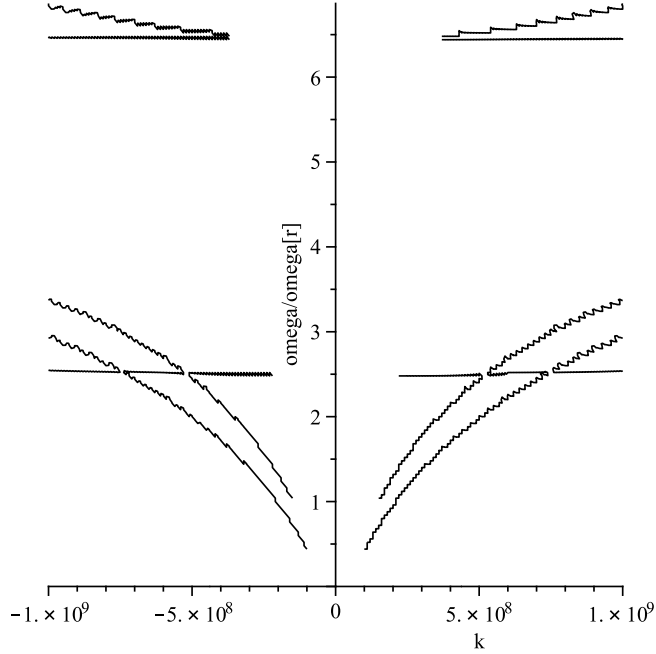}}}
    \caption{Non relativistic dispersion relation, for; $n=10^{20}cm^{-3}$, l=200 cm, $B_0=10^8$ G.}\label{11}
\end{figure}
One can clearly see the Langmuir and electromagnetic parts in
\textbf{Fig 2}. For better viewing we have changed the limiting
value in the MAPLE program to obtain \textbf{Fig 3}.
\begin{figure}[hbt]
  \centerline{\epsfxsize=3.5in \epsfysize=3.5in{\epsffile{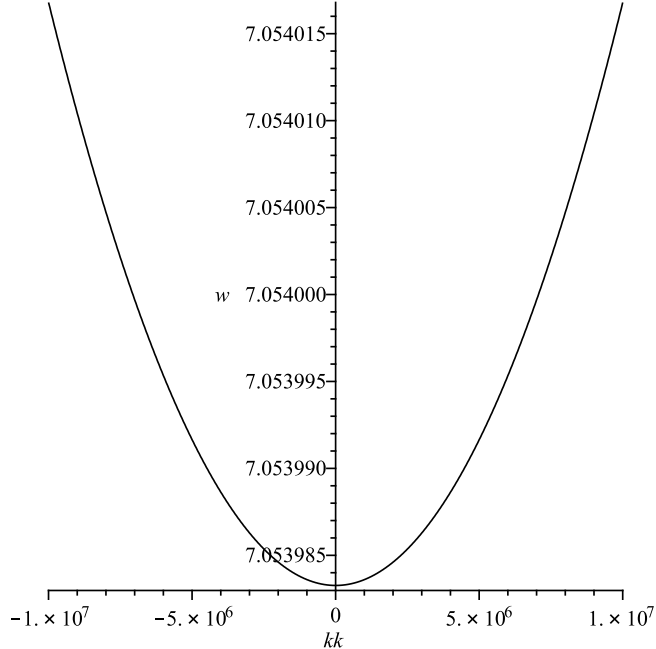}}}
    \caption{Non relativistic dispersion relation. The electromagnetic part of Fig 2.}\label{11}
\end{figure}
As we can see in \textbf{Fig 3}, the cutoff frequency of
electromagnetic wave is: \b\omega_{0}=7.053\omega_{r}.\e
\begin{figure}[hbt]
  \centerline{\epsfxsize=3.5in \epsfysize=3.5in{\epsffile{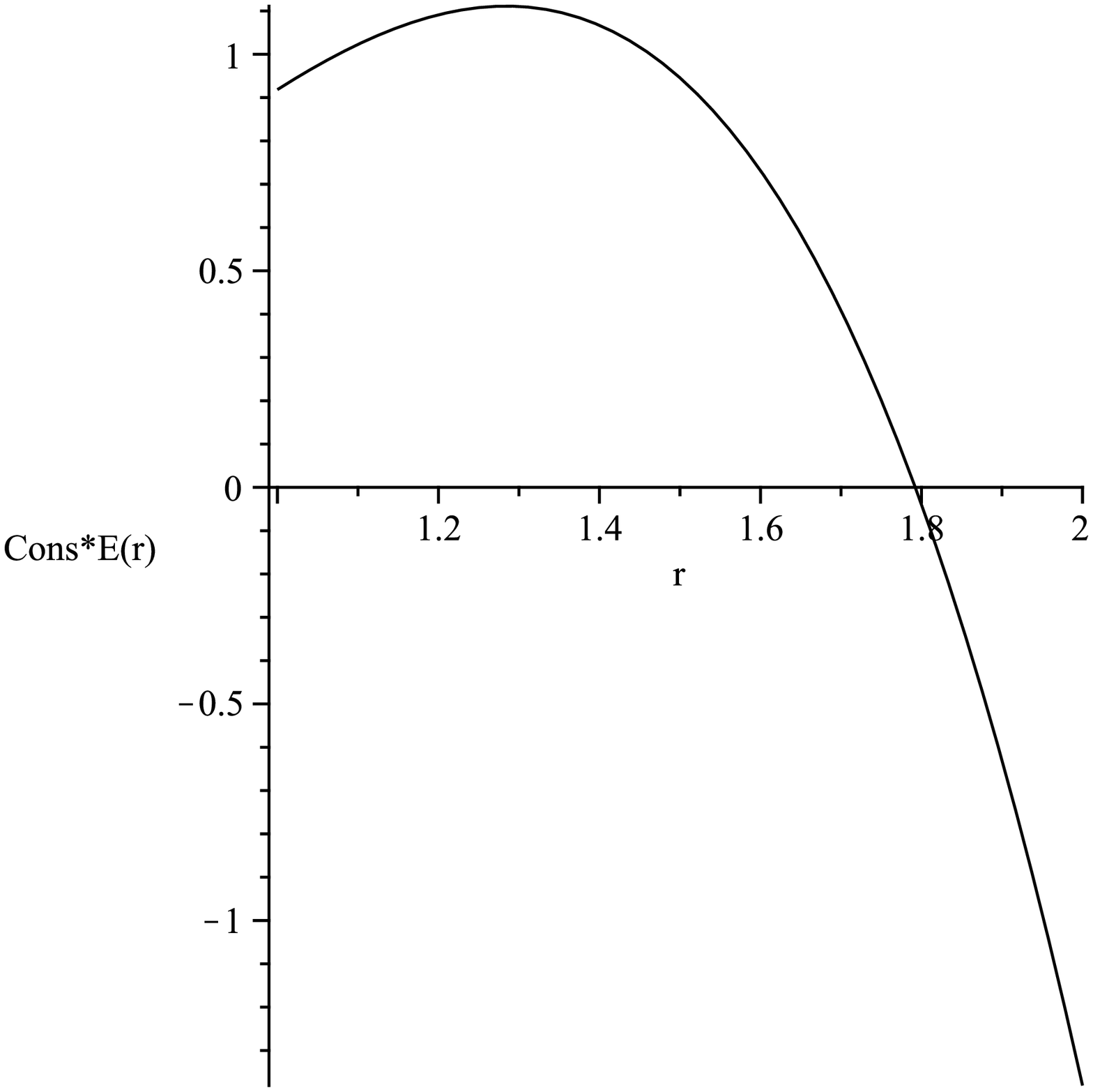}}}
    \caption{$E_{\phi}$ variation with r, for; $n=10^{20}cm^{-3}$, l=200 cm, $B_0=10^8$ G, and diameter of pipe is $r=1.8 cm$.}\label{11}
\end{figure}

In this study we will suppose only transverse wave exist. This
assumption is based on the fact that the shear tensor create
transversal motion. Due to viscosity of the electronic fluid,
transversal motion will create longitudinal motion. We will also
suppose that the electronic fluid is very dilute and hence there
is not any longitudinal motion. Thus to first order, gravitational
wave will create a consistent transversal mode. Note that we could
use proper dielectric tensor:
\b\epsilon=\left(
              \begin{array}{ccc}
                1 & 0 & 0 \\
                0 & 1 & 0 \\
                0 & 0 & 1-\omega_{p}^{2}/\omega^{2}\\
              \end{array}
            \right).
 \e
In the magnetized non neutral electronic plasma confined in a
metallic cylinder, the dielectric tensor is given by this matrix.
With this dielectric tensor we arrive at dispersion relation
\cite{Krall}:
\b(ka)^2=\frac{\omega^2a^2}{c^2}-\frac{p_{n\nu}^2}{1-\omega_{p}^2/\omega^2},\e
where $a$ is cylinder radius and $p_{n\nu}$  is $\nu$th root of
first type Bessel function of order n. This dispersion relation
has a cutoff frequency that is in full agreement with our previous
result Eq(32). Since we want to extend the problem to the
relativistic case, we have not used dielectric tensor. In our
consideration both Maxwell and momentum Equations have been
changed. To compute electric field $E_{\phi}$, we first obtain $k$
from $k=\frac{2\pi n}{L}$, where $L$ is  the length of the pipe
and $n$ is an integer number. With a specified $k$, one obtains
$\omega$ from dispersion relation and then electric field can be
obtained. The result is sketched in \textbf{Fig 4}.

We must note that $E_{\phi}$ is the solution of a homogenous
differential equation and hence it can be specified up to a
multiplication factor.

\section{3+1 formalism}

Although relativity joins space and time but sometimes splitting
of space and time is useful. We usually like to think physical
events as spacial events in different times. This is the 3+1
formalism; 3 for spatial dimensions and 1 for time dimension
\cite{Thorne2,Gourgoulhom,Elli1,Elli2,York,Poisson,Chop}. In 3+1
formalism, every physical quantity splits to two components, one
in the space like hypersurface and the other in the orthogonal
direction to it. Energy momentum tensor in this formalism is
\cite{Elli2,Landu}: \b T_{ab}^{dust} =\mu
u_au_b+q_au_b+u_aq_b+p\gamma _{ab}+\pi_{ab}, \e where $u_a$ is the
dust velocity component that have filled the whole spacetime
manifold, $\mu$ energy density, $q$ momentum density , $p$
isotropic pressure, $\pi$ traceless anisotropic pressure, and
$\gamma$ is projection operator. In this formalism, energy
momentum tensor of Maxwell's field is
\cite{Elli2,Tsaggas1,Thorne1} \b  T_{ab}^{(em)}
=\frac{1}{2}(E^2+H^2)U_aU_b+\frac{1}{6}(E^2+H^2)\gamma
_{ab}+2Q_{(a}U_{b)}+P_{ab}, \e where $~~~~E^2=E_a E^a$,$~~~H^2=H_a
H^a$,$~~~~Q_a=\eta_{abc}E^b H^C$ Poynting vector, and $P_{ab}$
traceless tensor: $$ P_{ab}=P_{\langle ab\rangle}=
\frac{1}{2}(E^{2}+H^2)\gamma _{ab}-E_aE_b-H_aH_b .$$  $E_a$ and
$H_a$ are the electromagnetic field components that can be
obtained from Faraday tensor, $F_{ab}$, by \b E_a=F_{ab}u^b,\;\;
H_{a}=\frac{1}{2}\eta_{abc}F^{ab} . \e

Where $\eta_{abc}$ is levi-Civita tensor.
In 3+1 formalism, matter and energy will be specified by $\mu$,
$q^a$, $P$, $\pi^{ab}$, $E^a $, $H^a $, $Q_a$, and $P_{ab}$.
Spacetime structure will be specified by, $\dot{u}_a$ fluid
acceleration, $\theta $ expansion coefficient, $E^{ab}$ electric
Weyl tensor, and $H^{ab}$ magnetic Weyl tensor. These quantities
are defined as follows:\b \theta=\tilde{\nabla}_{a}U^{a},(\theta
=3H ),\;\; \sigma_{ab}= \tilde{\nabla}_{\langle a}U_{b\rangle}
,\;\;\omega_{ab}= \tilde{\nabla}_{[ a}U_{b]} $$
$$E_{ab}=c_{acbd}U^{c}U^{d} ,\;\;
H_{ab}=\frac{1}{2}\eta_{ade}c_{bc}^{de}U^{c}, \e where
$\tilde{\nabla}$ is projection of covariant derivative in
hypersurface, $c_{abcd}$ the Weyl conformal curvature tensor, and H the Hubble constant. If $\omega^{ab}=0$, $\tilde{\nabla}$ will then equal
to 3 dimensional covariant derivative in the hypersurface, in such
case we denote it with $\nabla$. Dynamic of energy-matter
quantities and geometrical quantities can be obtained by
projection of Einstein equation along normal vector (orthogonal to
hypersurface). We list some of these equations that will be needed
later \cite{Elli2}: \b \dot{\mu}+\tilde{\nabla}_a q^a =-\theta
(\mu+p)-2(\dot{U}_a q^a )-(\sigma_b^a \pi _a ^b ),~~~~~~~~(\mu
~propagation~ Eq.)$$ $$ \dot{q}^{\langle
a\rangle}+\tilde{\nabla}^a p +\tilde{\nabla}_b \pi ^{ab}
 =\frac{-4}{3}\theta q^a -\sigma _b^a q^b-(\mu+p)\dot{U}^a-\dot{U}_b \pi^{ab}-\eta ^{abc}\omega_b q_c,~~~~~~~(q ~propagation~ Eq.)$$ $$ \dot{E}_{\langle
a\rangle}=(\sigma_{ab}+\eta_{abc}\omega^c-\frac{2}{3}\theta
\gamma_{ab})E^b +\eta_{abc}\dot{U}^b H^c+curl H_a-j_a ,~~~~~~~~(E_a~propagation~ Eq.)$$ $$ \dot{H}_{\langle a\rangle} =(\sigma_{ab}+\eta_{abc}\omega^c-\frac{2}{3}\theta \gamma_{ab})H^b+\eta_{abc}\dot{U}^bE^c
 -curl E_a ,~~~~~~~~(H_a ~propagation~ Eq.) $$ $$ \dot{\sigma}^{\langle ab\rangle}-\tilde{\nabla}^{\langle a}\dot{U}^{b\rangle}=
  \frac{-2}{3}\theta \sigma^{ab}+\dot{U}^{\langle a}\dot{U}^{b\rangle}-\sigma_c^{\langle a} \sigma^{b\rangle c}
  -\omega^{\langle a} \omega^{b\rangle}-(E^{ab}-\frac{1}{2}\pi^{ab}),~~~~~~(\sigma_{ab} ~propagation ~Eq.) $$ $$  (\dot{E}^{\langle ab\rangle}+\frac{1}{2}\dot{\pi}^{\langle
ab\rangle})-(curl H)^{ab}+\frac{1}{2}\tilde{\nabla}^{\langle
a}q^{b\rangle}
=\frac{-1}{2}(\mu+p)\sigma^{ab}-\theta(E^{ab}+\frac{1}{6}\pi^{ab})+3\sigma_c^{\langle
a}(E^{b\rangle c} -\frac{1}{6}\pi^{b\rangle c})$$
$$-\dot{U}^{\langle a}q^{b\rangle}+\eta^{cd\langle a}[2\dot{U}_c
H_d^{b\rangle}+\omega_c(E_d^{b\rangle}+\frac{1}{2}\pi_d^{b\rangle})],~~~~~~~~(E_{ab} ~propagation ~Eq.)$$ $$ \dot{H}^{\langle ab\rangle}+(curl E)^{ab}-\frac{1}{2}(curl \pi)^{ab}
 =-\theta H^{ab}+3\sigma_c^{\langle a}H^{a\rangle c}+\frac{3}{2}\omega^{\langle a}q^{b\rangle}
 -\eta^{cd\langle a}[2\dot{U}E_d^{b\rangle }$$ $$-\frac{1}{2}\sigma_c^{b\rangle}q_d-\omega_cH_d^{b\rangle}] ,~~~~~~~(H_{ab} ~propagation~Eq.) \e
where bracket in tensor or vector means as:
 $$  v^{\langle a\rangle}=\gamma_b^av^b\;\;\;\;\;\;
\&               \;\;\;\;\;\;  T^{\langle ab\rangle}=[\gamma_c^{(
a}\gamma_d^{b)}-\frac{1}{3}\gamma^{ab}\gamma_{cd}]T^{cd}. $$

In deriving Eqs(39), we suppose that dust tensor prevail to
electromagnetic tensor, also we have ignored gravitational self
interaction. In 3+1 formalism, $E^{ab}$ and $H^{ab}$ represent
gravitational wave and $\sigma^{ab}$ entered as an intermediate
field \cite {Maartens1,Elli3,Challinor,Elli4}. In Minkowskian
space $\omega=\theta=\dot{u}=\sigma=0$ and Eqs(39) reduce to
 \b   \frac{\partial n}{\partial t}+\vec \nabla .(n\vec v)=0,  $$
 $$ (\frac{\partial \vec v}{\partial t}+(\vec v.\vec \nabla )\vec v)^a=-\sigma_b^a-\frac{e}{m}
 (\vec E+\frac{\vec v\times \vec B}{c})^a ,  $$
 $$(\vec \nabla \times \vec B)_a=\frac{-i\omega}{c}E_a+\frac{4\pi}{c}(-nq\vec v)_a-\sigma_{ab}E^b, $$
  $$(\vec \nabla \times \vec E)_a=\frac{i\omega}{c}B_a+\sigma_{ab}H^b.\e
Note that we have added Lorentz force to second Eq of Eqs(39) to
obtain the second Eq of Eqs(40) \cite{Marklund1}. As we can see
from Eqs(40), the gravitational waves itself have not been entered
in these Eqs, but correlated field $\sigma$ have been appeared in
these Eqs.

\section{ Gravitational waves effect on bounded cold electronic plasma }

We have seen in section 3 that, if for any reason the fields in
the cylinder is perturbed then the resulting fields will satisfy
certain homogenous equations. Now, if we expose gravitational
waves on the plasma, we will get a non-homogenous deferential
equations with gravitational wave term at the right hand side. We
observe from dispersion relation that at a certain frequency, the
motion of the electrons will resonate and can pass through the
potential barrier. We follow section 3 to obtain non-homogenous
equations. To do that we replace the leading equations of section
3 with Eqs(40). To solve the resulting equations, we focus on an
spacial case that $\sigma^{ab}$ is diagonal. With this assumption,
we write velocity field in term of electromagnetic field and
gravitational field, $\sigma$,:

 \b v_\phi=aE_r+brB_z+fE_\phi+gr\sigma ,\;\;v_r=AE_\phi+DE_r+SrB_z+Gr\sigma,   \e
where \b   a=\frac{-e\omega_r/m}{(k\omega_r-\omega)^2-\omega_r^2}
,\;\;b=\frac{-e\omega^2_r/mc}{(k\omega_r-\omega)^2-\omega_r^2}=\frac{\omega_r}{c}a
,\;\;
f=\frac{i(k\omega_r-\omega)/m}{(k\omega_r-\omega)^2-\omega_r^2}
,\;\;
g=\frac{i(k\omega_r-\omega)/e}{(k\omega_r-\omega)^2-\omega_r^2}\omega_r
$$ $$ A=\frac{-e}{m\omega_r}+\frac{(k\omega_r-\omega)^2}{\omega_rm[(k\omega_r-\omega)^2-\omega_r^2]} ,\;\;
 D=\frac{i(k\omega_r-\omega)e/m}{(k\omega_r-\omega)^2-\omega_r^2}
 ,\;\;
S=\frac{ie\omega_r(k\omega_r-\omega)/mc}{(k\omega_r-\omega)^2-\omega_r^2},$$
$$
G=[\frac{-e}{m\omega_r}+\frac{(k\omega_r-\omega)^2}{\omega_rm[(k\omega_r-\omega)^2
 -\omega_r^2]} ]\frac{m\omega_r}{e}
 =A\frac{m\omega_r}{e}.\e

Now if we use Eqs(41) for the velocity field we can express
magnetic field in term of electric field by the Ampere Eqs(40):
\b(\frac{ik}{r}-S')B_z-ik'B_\phi=(\frac{-i\omega}{c}+D)E_r+A'E_\phi
 +G'r\sigma ,$$ $$-\frac{\partial B_z}{\partial r}-b'rB_z-S''\frac{\partial}{\partial r}(r^2 B_z)
 = (\frac{-i\omega}{c}+f')E_\phi+A''\frac{\partial}{\partial
r}(rE_\phi)+a'E_r
 +D''\frac{\partial}{\partial r}(rE_r)+g'r\sigma+G''\frac{\partial}{\partial r}(r^2\sigma),\e
where the primed and double primed quantity is defined as follow
\b  A'=\frac{-4\pi e n_0}{c}A ,\;\; a'=(\frac{-4\pi e
n_0}{c})(1+\frac{k\omega_r}{(\omega-k\omega_r)})a ,\;\;
A''=\frac{-4\pi e \omega_r n_0}{ic(\omega-k\omega_r)}A .\e If we
put Eq(43) in Faraday Eq(40) and obtain $E_\varphi$ from it, we
will obtain the counterpart of Eq(27): \b
(\frac{i\omega}{c}+f')E_\phi+A''\frac{\partial}{\partial
r}(rE_\phi)+a'M(r)\frac{\partial}{\partial
r}(rE_\phi)-a'N(r)E_\phi $$ $$ =\frac{-c}{i\omega+\sigma
c}\frac{\partial}{\partial r}(\frac{1}{r}\frac{\partial}{\partial
r}(rE_\phi))+ \frac{i kc}{i\omega+\sigma
c}\frac{\partial}{\partial
r}(\frac{M(r)}{r}\frac{\partial}{\partial r}(rE_\phi))
-\frac{ikc}{i\omega+\sigma c}\frac{\partial}{\partial
r}(\frac{N(r)}{r}E_\phi)-\frac{b'c}{i\omega+\sigma
c}\frac{\partial}{\partial r}(rE_\phi)$$ $$
+\frac{ikb'c}{i\omega+\sigma c}M(r) \frac{\partial}{\partial
r}(rE_\phi) -\frac{ikb'c}{i\omega+\sigma
c}N(r)E_\phi-\frac{S''c}{i\omega+\sigma c}\frac{\partial}{\partial
r}(r\frac{\partial}{\partial r}(rE_\phi))$$ $$
+(g'+2G'')r\sigma-\frac{ikc\sigma}{i\omega+\sigma
c}\frac{\partial}{\partial
r}(\frac{L(r)}{r})-\frac{ikb'c\sigma}{i\omega+\sigma c}L(r),\e
where M(r), N(r) and L(r) are defined as follows: \b
M(r)=(\frac{ik}{r}-S'r)(\frac{c}{r(i\omega+\sigma c)})/R(r) .\;\;
N(r)=A'/R(r) ,\;\; L(r)=G'r/R(r) ,$$ $$
R(r)\equiv(\frac{-i\omega}{c}+D')+(\frac{ik}{r}-S'r)(\frac{c}{r(i\omega+\sigma
c)})+\frac{ik'^2c}{\omega}  \e

As what has been said in section 3, $E_{\phi(r)}$ can be obtained
from Eq(45) and the $E_{r}(r)$ can be obtained from following
equation: \b E_{r}(r)=M(r)\frac{\partial}{\partial
r}(rE_{\phi(r)})-N(r)E_{\phi}(r),\e and magnetic field is: \b
B_{\phi}=\frac{c k'}{\omega}E_{r},\e \b
B_{z}=\frac{c}{r(i\omega+\sigma c)}(\frac{\partial}{\partial
r}(rE_{\phi})-ikE_{r}).\e Velocity field are obtained from
Eqs(41). Kinetic energy of an electron near the far end of the
cylinder for $r=4$ is sketched in \textbf{Fig 5}. As we can see
the kinetic energy will increase with external gravitational
field.

\begin{figure}[hbt]
  \centerline{\epsfxsize=3.5in \epsfysize=3.5in{\epsffile{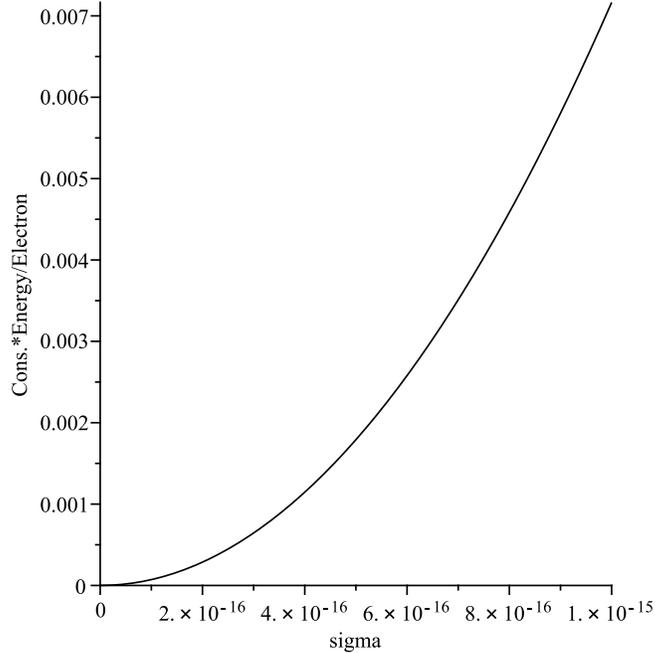}}}
    \caption{Kinetic energy of electronic plasma particle will be increased by GW, for; $n=10^{20}cm^{-3}$, l=200 cm, $B_0=10^8$ G.}\label{11}
\end{figure}

According to general relativity the geometry of spacetime is
determined by matter and energy. The motion of heavy mass in the
spacetime causes time dependent variation in its curvature that is
called GW. The effect of gravitation that entered in this study
appears as $\sigma$ field that are originated from the motion of
far heavy (binary) stars. Having the order of magnitude of
$\sigma$ that belongs to the possible sources of GW, one can
evaluate the kinetic energy of the electrons. If one knows the
system parameters like electron density, the size of the pipe, and
potential of the rings, he/she can in principle create such
appropriate conditions that the electrons inside the pipe can pass
through the potential barrier, that is a direct sign of existence
of GW. This is main idea of the present paper. With a proper
$\sigma$ field strength, a typical electron can get enough energy
to cross the potential barrier. This is a remarkable result
because based on this calculation one can hope to detect the GW
directly by the method described here.

\section{ Conclusions}

We have proposed an experimental sketch to detect gravitational
waves(GW) \emph{directly}, using an cold electronic plasma in a
long pipe. By considering an cold electronic plasma in a long
pipe, the Maxwell equations in 3+1 formalism have been invoked to
relate gravitational waves to the perturbations of plasma
particles. It has been shown that the impact of GW on cold
electronic plasma causes disturbances on the pathes of the
electrons. Then we have shown that those electrons that absorb
energy from GW will pass through the potential barrier at the end
of the pipe. Therefore, crossing of some electrons over the
barrier will imply the existence of the GW.

\vspace{0.5cm} \noindent {\bf{Acknowledgements}}: One of authors
(O.J.) would like to thanks to his colleague
 Dr. M. Khanpour for his useful discussion and kindly helps.
 
\end{document}